\documentclass[lettersize,10pt, journal]{IEEEtran}

\usepackage{amsmath,amsfonts}
\usepackage{algorithm}
\usepackage{algpseudocode}
\usepackage{array}
\usepackage{amssymb}
\usepackage{graphicx}
\usepackage[strings]{underscore}
\usepackage{cite}
\usepackage{xcolor}
\usepackage{colortbl,booktabs}
\usepackage{tabularx}
\usepackage{threeparttable}
\usepackage{multirow}
\usepackage{pifont}
\usepackage{enumitem}
\usepackage[numbered]{bookmark}
\usepackage{hyperref}
    \hypersetup{
        bookmarksnumbered=true,
        bookmarksopen=true,
        bookmarksopenlevel=5,
        colorlinks=true,
        citecolor=black,
        filecolor=black,
        linkcolor=black,
        urlcolor=black,
        pdfstartview=Fit,
        pdfpagemode=UseOutlines
    }

\definecolor{darkgreen}{RGB}{0,128,0}

\begin{document}
\title{DREAM-PCD: Deep Reconstruction and Enhancement of  mmWave Radar Pointcloud}

\author{Ruixu~Geng,
        Yadong~Li,
        Dongheng~Zhang,
        Jincheng~Wu,
        Yating~Gao,
        Yang~Hu,
        Yan~Chen,~\IEEEmembership{Senior~Member,~IEEE}

\thanks{This work has been submitted to the IEEE for possible publication. Copyright may be transferred without notice, after which this version may no longer be accessible.
}
\thanks{Ruixu Geng, Yadong Li, Dongheng Zhang, Jincheng Wu, Yating Gao, Yan Chen are with the School of Cyber Science and Technology, University of Science and Technology of China, Hefei 230026, China (e-mail: eecyan@ustc.edu.cn)}
\thanks{Yang Hu is with the School of Information Science and Technology, University of Science and Technology of China, Hefei 230026, China (e-mail: eeyhu@ustc.edu.cn)}

}

\markboth{IEEE TRANSACTIONS ON IMAGE PROCESSING, VOL. XX, 202X}
{XX \MakeLowercase{\textit{et al.}}: DREAM-PCD: Deep Reconstruction and Enhancement of  mmWave Radar Pointcloud}

\maketitle

\begin{abstract}
    Millimeter-wave (mmWave) radar pointcloud offers attractive potential for 3D sensing, thanks to its robustness in challenging conditions such as smoke and low illumination. 
    However, existing methods failed to simultaneously address the three main challenges in mmWave radar pointcloud reconstruction: specular information lost, low angular resolution, and strong interference and noise. 
    In this paper, we propose DREAM-PCD, a novel framework that combines signal processing and deep learning methods into three well-designed components to tackle all three challenges: Non-Coherent Accumulation for dense points, Synthetic Aperture Accumulation for improved angular resolution, and Real-Denoise Multiframe network for noise and interference removal. 
    Moreover, the causal multiframe and ``real-denoise" mechanisms in DREAM-PCD significantly enhance the generalization performance. 
    We also introduce RadarEyes, the largest mmWave indoor dataset with over 1,000,000 frames, featuring a unique design incorporating two orthogonal single-chip radars, lidar, and camera, enriching dataset diversity and applications. 
    Experimental results demonstrate that DREAM-PCD surpasses existing methods in reconstruction quality, and exhibits superior generalization and real-time capabilities, enabling high-quality real-time reconstruction of radar pointcloud under various parameters and scenarios. 
    We believe that DREAM-PCD, along with the RadarEyes dataset, will significantly advance mmWave radar perception in future real-world applications.
\end{abstract} 


\IEEEpeerreviewmaketitle

\section{Introduction} \label{sec:intro}

Perceiving 3D environments is critical for robotics, autonomous vehicles, and unmanned aerial systems~\cite{gutmann20083Dperceptionandenvironmentmapgeneration}. 
Among different representations such as voxel~\cite{deng2021voxelrcnn}, 2D projection~\cite{zhou2018voxelnet}, pointcloud~\cite{rusu2011PCL} and mesh~\cite{hu2021sailvos3d}, pointcloud has become the leading representation for 3D perception tasks due to its memory efficiency and its capacity to flexibly and accurately represent objects of varying shapes and sizes~\cite{pomerleau2015PCDregistrationreview}, driving progress in subfields like 3D object detection~\cite{shi2020pointGNN,lang2019pointpillars}, tracking~\cite{schulman2013trackingdeformableinPCD,qi2020p2bPCDtracking}, and semantic segmentation~\cite{landrieu2018largescalepcdsegmentic,wang2019graphattentionforpcdsegmentic}.
However, vision-based pointcloud generation process, including LiDAR and stereo imaging, face limitations such as eye safety~\cite{roriz2021automotiveLidarSurvey}, privacy concerns~\cite{chen2018context}, high cost, and vulnerability~\cite{vargas2021overviewofAutonomousVehiclesSensors} to extreme weather, hindering their widespread adoption.
To enable ubiquitous 3D sensing, the pointcloud generation process must satisfy (1) high environmental robustness, (2) low cost, and (3) no health or privacy threats.

\begin{figure}[t!]
   \centering
   \includegraphics[width=0.9\linewidth]{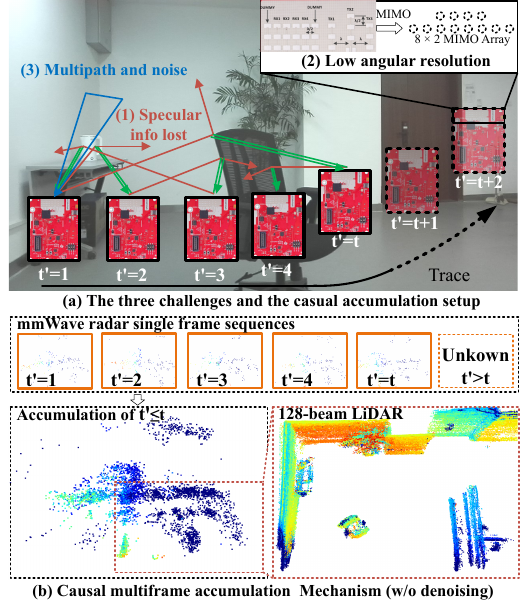}
   \caption{Causal multiframe accumulation mechanism for mmWave pointcloud reconstruction. (a) The three main challenges in mmWave pointcloud reconstruction. The idea of multiframe accumulation can mitigate the low angular resolution and sparsity issues. (b) Experiments show that single-frame radar pointcloud are extremely sparse, while accumulated multiframe pointcloud significantly enrich the scene features. Furthermore, the application of causal systems helps maintain real-time performance without compromising accuracy.}
   \label{fig:fourchallenges}
\end{figure}

Fortunately, mmWave radar pointcloud, produced by low-cost Frequency Modulated Continuous Wave (FMCW) radar, meets the three requirements for 3D sensing~\cite{guan2020throughFogHRRadarImagingCVPR,yadongGestureTMC,binbinUnsupervised} and achieves 4D pointcloud generation using LiDAR-like distance resolution, MIMO technology for angle estimation, and velocity perception through Doppler estimation~\cite{zhou2022towardsDeepRadarPerceptionForAutonomousDriving}.
Consequently, mmWave radar pointcloud represents a valuable and indispensable solution in future 3D perception systems~\cite{prabhakara2022RadarHD,cheng2022RPDNet,qian20203millipoint,cai2023millipcd,jiang20234dHighResolutionImageryofPCDFormmWaveRadar}.
However, current mmWave radar pointcloud lacks important echo information due to specular reflection, have low angular resolution, and suffer from multipath interference and noise~\cite{sun2021deeppoint,sun20213drimr}. 
As shown in Figure~\ref{fig:fourchallenges}-(a), these factors introduce three main challenges in mmWave radar pointcloud.

\vspace{0.2em}
\noindent \textbf{Challenge 1: Specular information lost and sparsity.} 
The 77 GHz mmWave radar, with a wavelength of approximately 4 mm, exhibits specular reflection as the predominant factor in the Bidirectional Reflectance Distribution Function (BRDF) on most object surfaces~\cite{xing2019indoormmWavechannelProperties}. This results in the inability to receive a significant portion of reflected signals, causing information loss and sparse pointcloud, as shown in Figure~\ref{fig:fourchallenges}-(a).

Multi-frame stacking from different views serves as a fundamental solution to tackle sparsity issues, benefiting from recent breakthroughs in odometry that enable cm-level location estimation. However, comprehensive research on multi-frame stacking for mmWave radar pointcloud reconstruction remains \textit{scarce}~\cite{prabhakara2022RadarHD}.
Additionally, multi-frame stacking introduces \textit{significant noise accumulation}.

\vspace{0.2em}
\noindent \textbf{Challenge 2: Low angular resolution.}
Low-cost mmWave radar devices have limited antenna numbers, resulting in fewer spatial spectrum sampling points and lower angular resolution~\cite{qian20203millipoint}.

To address this challenge, a common strategy is increasing the radar aperture size by employing large-scale antenna arrays or Synthetic Aperture Radar (SAR) principle~\cite{oliver1989syntheticApertureRadar,PuWei2021DeepSARImaging}. However, large arrays introduce size and cost constraints~\cite{haupt2015antennaArryDevelopments}, while SAR requires \textit{precise position estimation or regular motion patterns} (e.g., milliPoint~\cite{qian20203millipoint}) for coherent phase combination. In complex indoor environments, maintaining precise position estimation over long distances and varying trajectories is difficult~\cite{cai2023millipcd}.

\vspace{0.2em}
\noindent \textbf{Challenge 3: Severe interference.} 
The mmWave radar pointcloud also contains severe multipath and noise interference. Among them, multipath effects arise due to the unique propagation characteristics of mmWave radar signals in complex environments, including specular reflection and diffraction. Noise originates from factors such as phase errors and misjudgments in point detection. These interferences are particularly pronounced in ubiquitous 3D sensing scenarios, especially in indoor environments, giving rise to ``ghost points" and substantially degrading pointcloud quality~\cite{guan2020throughFogHRRadarImagingCVPR}.

Since the physical removal of multipath and noise interference is highly challenging~\cite{gennarelli2014multipathGostsRadarImaging,xiong2013arraytrack}, existing methods~\cite{sun2021deeppoint,sun2021deeppoint,sun20213drimr,cheng2022RPDNet,prabhakara2022RadarHD} typically employ deep learning to establish a mapping between radar signals and LiDAR pointcloud. However, these methods are prone to \textit{overfitting} due to the limited information contained in radar signals~\cite{li2023azimuthsuperresolution,jiang20234dHighResolutionImageryofPCDFormmWaveRadar} and the significant differences and ill-posedness between radar and LiDAR data.

\vspace{0.2em}
\noindent \textbf{Limitation: Dataset.}
Learning-based reconstruction approaches heavily rely on large-scale and high-quailty datasets to ensure the model's robustness.
However, existing public radar pointcloud datasets suffer from lacking raw ADC data for comprehensive research~\cite{zhou2022towardsDeepRadarPerceptionForAutonomousDriving,caesar2020nuscenes,schumann2021radarscenes,bansal2020pointillism}, limited diversity in indoor environments for robust representation learning~\cite{lim2021Radical,mostajabi2020Zendar}, and the presence of co-frequency interference~\cite{kramer2022coloradar}.

\vspace{0.2em}
\noindent \textbf{Our approach.} 
This paper presents DREAM-PCD, an innovative framework designed for generating high-quality pointcloud using low-cost radar. The core idea behind DREAM-PCD is simple yet effective: combining the interpretability of traditional signal processing methods with the representational power of data-driven network to simultaneously address the aforementioned challenges.
Our approach comprises three key components: the traditional (1) Non-Coherent Accumulation (NCA) and (2) Synthetic Aperture Accumulation (SAA) that address specular information lost from multiple views and overcome low resolution using synthetic aperture, respectively, and the data-driven (3) Real-Denoise Multiframe (RDM) network for mitigating noise and multipath interference with a novel ``Real-Denoise" strategy. These components together form the DREAM-PCD framework. Moreover, DREAM-PCD leverages causal multiframe accumulation mechanism (as shown in Figure~\ref{fig:fourchallenges}-(b)), utilizing scene priors while maintaining real-time performance.To address the challenge of dataset scarcity, we introduce RadarEyes, a high-quality large-scale indoor radar dataset equipped with two orthogonal single-chip radars, one LiDAR, and one camera. In contrast to previous studies~\cite{prabhakara2022RadarHD,cheng2022RPDNet,qian20203millipoint} that addressed only one or two challenges, DREAM-PCD tackles all three main challenges, outperforming other methods and enabling high-quality radar pointcloud generation using low-cost single-chip radar. Our major contributions are summarized as follows.

\vspace{0.1em}
\noindent (1) We introduce DREAM-PCD, a novel mmWave pointcloud reconstruction framework, specifically designed for ubiquitous environments using low-cost 3T4R radar. To the best of our knowledge, this comprehensive framework, encompassing three components (NCA, SAA, and RDM Network), is the first attempt to simultaneously address the three main challenges in mmWave radar pointcloud reconstruction.

\vspace{0.1em}
\noindent (2) We propose the Real-Denoise Multiframe (RDM) network to tackle the overfitting issue in deep learning-based mmWave pointcloud generation tasks. RDM transforms the pointcloud reconstruction task into a point-wise binary classification problem (i.e., real-denoise instead of regression to LiDAR points) by designing a new ground truth, significantly reducing the ill-conditioning and enhancing its generalization ability.

\vspace{0.1em}
\noindent (3) We introduce RadarEyes, a comprehensive large-scale dataset with four aligned sensors for mmWave radar pointcloud generation. Containing over 1,000,000 data frames from 300 indoor and 10 outdoor scenes, RadarEyes significantly surpasses the previous largest indoor dataset, Coloradar, which has only 121,830 frames. To our knowledge, this makes RadarEyes the largest indoor mmWave radar dataset available.

\vspace{0.1em}
\noindent (4) Extensive experiments demonstrate the feasibility of generating high-density pointcloud with 3T4R mmWave radar in various environments, with DREAM-PCD outperforming existing methods in reconstruction quailty~\cite{cheng2022RPDNet,prabhakara2022RadarHD} and offering superior generalization capability. \textit{DREAM-PCD framework and RadarEyes dataset will be made publicly available
in \url{https://sites.google.com/view/dream-pcd} upon paper acceptance.}

\section{Related Work}

\noindent \textbf{mmWave Radar Pointcloud Generation}~\cite{waldschmidt2021automotiveRadarFromFirstEffortsToFutureSystems} 
faces three main challenges including low angular resolution, sparse pointcloud, and severe interference due to antenna and propagation properties. To increase angular resolution, large antenna apertures and Synthetic Aperture Radar (SAR) are common approaches, while the former is expensive and less portable~\cite{haupt2015antennaArryDevelopments} and SAR demands precise position estimation, limited radar motion and imaging range~\cite{sengupta2020reviewOfRecentAdvancementsIncludingMachineLearningOnSARUsingmmWaveRadar}, e.g., MilliPCD~\cite{cai2023millipcd} and MilliPoint~\cite{qian20203millipoint}.
Multi-frame accumulation is a widely-used strategy to address sparsity and information loss in other tasks like 2D RadarHD~\cite{prabhakara2022RadarHD} and human sensing~\cite{CapturingHumanFigureMultiView}. Nonetheless, in the 3D mmWave radar pointcloud domain, rare work has attempted this approach. Deep learning, owing to its powerful representation capabilities, is a favored method to counter severe interference. DeepPoint~\cite{sun2021deeppoint} and 3DRIMR~\cite{sun20213drimr} demonstrate potential by learning pointcloud of basic objects, whereas RPDNet~\cite{cheng2022RPDNet}, ADC SR~\cite{li2023azimuthsuperresolution} and CV-DCN~\cite{jiang20234dHighResolutionImageryofPCDFormmWaveRadar} utilizes deep learning across various pointcloud generation stages. Despite their potential, these methods struggle with overfitting and insufficient input information. Up to date, no existing work, aside from DREAM-PCD, has concurrently addressed these three challenges, thereby achieving superior pointcloud generation results even in complex indoor environments.

\vspace{0.1em}
\noindent \textbf{Pointcloud Denoising} 
aims to remove noise from acquired pointcloud to produce cleaner ones. Existing LiDAR pointcloud denoising networks typically leverage rich input features, using various architectures like convolution-based~\cite{li2018pointcnn,wu2019pointconv}, graph-based~\cite{wang2019DGCNN,TIP2021DynamicPCDDenoising}, generative model-based~\cite{wang2017shapeinpaitingGAN}, or transformer based~\cite{pan2021pointformer,zhao2021pointtransformer} to simultaneously perform denoising and regression (from wrong $x,y,z$ to correct $x,y,z$) tasks~\cite{Luo2021ICCVScoreBasedPCDDenoising,HermosillaTotalDenoising}, resulting in high-quality pointcloud outputs. However, for mmWave radar supervised by LiDAR, regression is highly challenging and ill-posed. More explicitly, using LiDAR data as groundtruth while networks that only input mmWave radar data would lead to overfitting, such as RadarHD~\cite{prabhakara2022RadarHD} and milliPCD~\cite{cai2023millipcd}. DREAM-PCD firstly proposed the ``real-denoise" mechanism by utilizing denoised mmWave radar pointcloud for supervision, enabling the network to focus on the denoising task and learn noise characteristics rather than relying on scene priors, thereby achieving high generalization capabilities.

\vspace{0.1em}
\noindent \textbf{Radar Datasets}
have been the focus of several efforts in recent years, leading to the creation of mmWave radar datasets such as nuScenes~\cite{caesar2020nuscenes}, RadarScenes~\cite{schumann2021radarscenes}, RaDICaL~\cite{lim2021Radical}, and Zendar~\cite{mostajabi2020Zendar}. Regrettably, there is still a limited number of mmWave radar datasets available for both indoor and outdoor environments, as well as including raw ADC data~\cite{zhou2022towardsDeepRadarPerceptionForAutonomousDriving}. To the best of our knowledge, the proposed RadarEyes is the largest indoor mmWave radar dataset to date, containing raw ADC data, diverse indoor scenes, and high-quality multimodal data.

\section{Background}

In this section, we provide a brief overview of the conventional process for generating mmWave radar pointcloud, followed by an introduction to deep learning-based mmWave PCD generation methods.

\subsection{mmWave PCD Generation Pipeline} \label{sec:radarModel}
FMCW technology is widely adopted by existing Commercial-Off-The-Shelf (COTS) mmWave radars 
to generate 4D mmWave radar pointcloud including the distance distance $r$, velocity $v$, azimuth angle $\psi$, and elevation angle $\theta$.
In the following, we briefly present the theoretical foundation of FMCW radars to ease subsequent anlysis and would refer the readers to~\cite{rao2017introductionTIFMCWradar} for a detailed pointcloud generation process.

\vspace{0.4em}
\noindent \textbf{Range estimation.} 
The range estimation of FMCW radars is achieved by mixing and low-pass filtering the received signal (i.e., the raw ADC data), utilizing the relationship between intermediate frequency $\Delta f$ and distance $d = \frac{c}{2k_f} \Delta f$. The distance resolution is given by $d_{res} = \frac{c}{2B}$, where $c$ represents the speed of light, $k_f$ is the frequency sweep rate, and $B$ represents the bandwidth. For instance, a commercial mmWave radar system operating in the 77-81 GHz has a distance resolution of approximately 4 cm. The maximum detection range can be expressed as $d_{max} = \frac{F_s c}{2 k_f}$, where $F_s$ denotes the sampling rate.

\vspace{0.4em}
\noindent \textbf{Velocity estimation.} 
Additionally, FMCW signals can provide velocity information via the Doppler effect exhibited in adjacent chirps, represented by $v = \frac{\lambda f_d}{2 T_c}$, where $f_d$ is the Doppler frequency shift, $T_c$ denotes the time interval between two consecutive chirps, and $\lambda$ refers to the wavelength. According to signal sampling theory, the velocity resolution of FMCW radar is given by $v_{\text{res}} = \frac{\lambda}{2T_f}$, and the maximum measurable velocity is $v_{\text{max}} = \frac{\lambda}{4 T_c}$. Here, $T_f$ represents the duration of each radar frame.

\noindent \textbf{Azimuth and elevation angle estimation.} 
For MIMO radar arrays, the target's azimuth and elevation angles can be determined by measuring the phase difference between adjacent antennas. Considering a uniform linear array with $N$ antennas and an antenna spacing of $d_a$, the received signal phase difference, denoted as $\Delta \phi$, is related to the azimuth angle $\theta$ by the following equation: $\Delta \phi = \frac{2\pi d_a}{\lambda} \sin{\theta}$. Thus, the azimuth angle $\theta$ can be obtained as $\theta = \arcsin{\frac{\Delta \phi \lambda}{2\pi d_a}}$. The maximum detectable angle is given by $\theta_{max} = \arcsin{\frac{\lambda}{2d_a}}$, while the angle resolution is expressed as 
\begin{equation} \label{eq:thetaRes}
   \theta_{res} = \frac{\lambda}{Nd_a \cos \theta}.
\end{equation}
For widely-used single-chip commercial radars (e.g., TI 1843), the synthesized antenna array is a sparse $8\times 2$ array, 
leading to an azimuth angle resolution of $14.3^\circ$ and an elevation angle resolution of $57.29^\circ$, both of which are significantly lower than that of 128-beam LiDAR systems ($\leq 0.1^\circ$).

\vspace{0.4em}
\noindent \textbf{Detector and PCD Generation.} 
To generate mmWave radar pointcloud, continuous-representation radar imaging is first achieved using the above distance, velocity, and angle estimations. Subsequently, a target detector, such as the commonly used Cell Averaging Constant False Alarm Rate Detector (CA-CFAR), is required to obtain a discrete 3D scene representation. CA-CFAR estimates noise power $P_{\text{noise}}$ and threshold $T$ from surrounding cells in Range-Doppler maps as $T = \alpha \cdot P_{\text{noise}} = \alpha \cdot \frac{1}{N} \sum_{m=1}^{N} x_m$, where $\alpha$ denotes the threshold factor and $x_m$ is the sample in each training cell~\cite{gandhi1988analysisofCFAR}.
After CFAR detection, each detected $(r^i, v^i)$ corresponds to an $N_{RX} \times N_{TX}$ signal matrix. MIMO virtual antenna synthesis and azimuth-elevation angle FFT are employed to yield angle maps, and the Prominence algorithm is utilized to extract peak points {$(\theta^{ij}, \phi^{ij}), j = 1,2, ..., J$}. Finally, the mmWave radar pointcloud is generated by transforming coordinates from $(r,\theta,\phi,v)$ to $(x,y,z,v)$~\cite{kramer2022coloradar}.

\vspace{0.4em}
\noindent \textbf{Analysis of Challenges in mmWave PCD Generation.} Based on the background knowledge, we can briefly analyze the three challenges mentioned in Section~\ref{sec:intro}: information lost, low angular resolution and severe interference. (1) \textit{Information lost} arises from two aspects: unreceived target echoes due to specular reflections, and low-reflectivity target points ignored by the detector. (2) \textit{Low angular resolution} mainly results from the limited radar aperture size, as illustrated by Eq.~\ref{eq:thetaRes}. (3) \textit{Severe interference} originates from inherent multipath effects introduced in the original ADC data and false alarms generated by the detector.
Physical-based traditional methods can alleviate low angular resolution (e.g., SAR~\cite{cai2023millipcd,qian20203millipoint}) and sparsity (superimposition at different locations), but they are limited in removing noise highly correlated with scene characteristics.

\subsection{Learning-based mmWave Radar pointcloud Reconstruction}
Unlike methods that reconstruct pointcloud using physical models, deep learning-based methods directly learn the mapping from radar data to high-quality pointcloud through a neural network:
\begin{equation}
   \mathcal{P}_{hq}=\Theta(F(Radar_{ADC})).
   \label{eq:deepPCD}
\end{equation}
Here, $Radar_{ADC}$ represents the input raw ADC data, and $\mathcal{P}_{hq}$ denotes the desired high-quality pointcloud. $F(Radar_{ADC})$ signifies the range-Doppler (RD) map, range-angle (RA) map, or low-quality pointcloud obtained after certain traditional signal processing steps. $\Theta$ represents the neural network with input $F(Radar_{ADC})$ and output $\mathcal{P}_{hq}$. Typically, most studies use LiDAR-acquired pointcloud $\mathcal{P}_{LiDAR}$ as the target to optimize network weights. As deep learning-based methods learn prior scene information, they achieve significantly better visual results than traditional methods. However, the mapping between $\mathcal{P}_{LiDAR}$ (about 300THz) and the network input $F(Radar_{ADC})$ (about 77GHz) is highly complex and ill-posed. Consequently, existing deep learning approaches often face challenges with generalization and limited applicability to unseen environments~\cite{prabhakara2022RadarHD}. Moreover, deep learning methods can neither inherently improve imaging resolution nor address scene information loss and sparsity caused by specular reflections~\cite{cheng2022RPDNet}.

Our proposed DREAM-PCD framework combines the advantages of both physics-based~\cite{qian20203millipoint} and learning-based methods~\cite{cheng2022RPDNet,prabhakara2022RadarHD,cai2023millipcd}, addressing all the three main challenges of low angular resolution, noise interference, and sparsity. 

\section{System Design} \label{sec:framework}

In this section, we first provide an overview of the proposed DREAM-PCD framework, followed by a detailed description.

\begin{figure*}[h]
    \centering
    \includegraphics[width=0.95\linewidth]{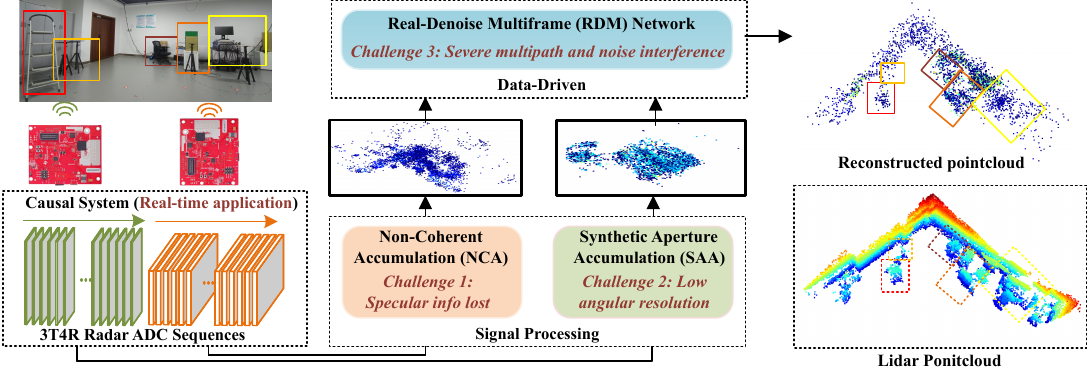}
    \caption{An overview of the proposed DREAM-PCD framework. The DREAM-PCD framework primarily consists of three components: (1) Multi-Frame Non-Coherent Accumulation (NCA), designed to alleviate the sparsity caused by specular reflections. (2) Synthetic Aperture Accumulation (SAA), employed to enhance angular resolution. (3) Real-Denoise Multiframe (RDM) network, dedicated to addressing multipath interference and noise. }
    \label{fig:DREAM-PCD}
\end{figure*}

\subsection{Overview of DREAM-PCD Framework}

This paper aims to combine the advantages of physics-based methods and deep learning-based methods to address all the three major challenges of mmWave radar pointcloud introduced in Section~\ref{sec:intro}, namely sparsity, low angular resolution, together with noisy and multipath interference. Hence, the proposed DREAM-PCD framework has three components: physics-based Non-Coherent Accumulation (NCA) to enhance pointcloud density, Synthetic Aperture Accumulation (SAA) to improve angular resolution; and deep learning-based Real-Denoise Multiframe (RDM) network to remove severe multipath interference and noise. Besides, the whole framework is designed as a causal system for real-time performance.

\vspace{0.4em}
\noindent \textbf{NCA} employs a multi-frame non-coherent accumulation method to alleviate information lost and sparsity caused by specular reflections. It merges multiple frames of mmWave radar data to obtain non-coherent multi-frame pointcloud $\mathcal{P}_{NCA}$, which encapsulates scene information from different viewpoints and mitigates sparsity and ill-posedness~\cite{cai2023millipcd}.

\vspace{0.4em}
\noindent \textbf{SAA} enhances angular resolution using coherent accumulation techniques. Employing the temporal back-projection (BP) algorithm~\cite{ulander2003SARBackProjection}, it increases antenna aperture through the field of view (FoV) filtering, resampling, and back-projection, generating a coherent multi-frame pointcloud, $\mathcal{P}_{SAA}$, with richer high-frequency information.
To satisfy the high frame rate requirement for synthetic aperture and acquire higher elevation resolution, an additional vertically placed radar (as shown in Figure~\ref{fig:DREAM-PCD}) is utilized for coherent data collection.
\vspace{0.4em}

\noindent \textbf{RDM network} aims to enhance pointcloud quality by learning the mapping between multi-frame mmWave radar and LiDAR pointcloud. It features two key innovations compared to existing approaches: (1) By taking $\mathcal{P}_{NCA}$ and $\mathcal{P}_{SAA}$ as inputs, the pointcloud reconstruction task is transformed into a more manageable ``real denoising task", which improves generalizability. (2) It learns the mapping between multi-frame mmWave radar data and LiDAR pointcloud, rather than relying on single-frame mappings, which allows for better capture of local and global scene features. These innovations enable the RDM network to produce high-quality and generalizable reconstruction results.

\vspace{0.4em}
\noindent \textbf{Causal system design.} DREAM-PCD is designed as a causal system, only accumulating radar data from previous time instances at time $t$ (as shown in Figure~\ref{fig:fourchallenges}-(b)). This ensures a real-time processing and output of high-quality pointcloud, which is suitable for autonomous navigation and other real-time applications.

The overview of the DREAM-PCD framework is illustrated in Figure~\ref{fig:DREAM-PCD}. It can be seen that, with an effective design, DREAM-PCD achieves high-quality pointcloud generation only using the low-resolution and cost-effective 3T4R mmWave radar.

\begin{algorithm}
    \caption{Non-Coherent Accumulation (NCA)}
    \label{alg:NCA}
    \begin{algorithmic}[1]
    \Require Input mmWave radar ADC data $D$
    \Ensure Output pointcloud $\mathcal{P}_{NCA}$
    
    \Function{SingleFramePointCloud}{$D$} (Sec.~\ref{sec:radarModel})
        \State Perform distance FFT and Doppler FFT on $D$ to obtain RDM
        \State Apply CFAR + 2D AOA estimation + peak detection
        \State \Return pointcloud for each frame $\mathcal{P}_{NCA, i}$
    \EndFunction
    
    \Function{CoordinateTrans}{$\mathcal{P}_{NCA, i}$, $t_i$}
        \State Obtain position $p_{x,i}, p_{y,i}, p_{z,i}$ and orientation $q_i$ at timestamp $t_i$
        \State Establish rotation transformation Q
        \State Apply coordinate transformation: $\mathcal{P}_{NCA} = \Sigma_{i=0}^{N} Q(P_{NCA, i}, q_{i}) + [p_{x,i}, p_{y,i}, p_{z,i}]$
        \State \Return output pointcloud $\mathcal{P}_{NCA}$
    \EndFunction
    
    \State $\mathcal{P}_{NCA, i} \leftarrow$ \Call{SingleFramePointCloud}{$D$}
    \State $\mathcal{P}_{NCA} \leftarrow$ \Call{CoordinateTrans}{$\mathcal{P}_{NCA, i}$, $t_i$}
    
    \end{algorithmic}
\end{algorithm}

\subsection{Non-Coherent Accumulation (NCA)}

The NCA fuses pointcloud frames from different viewpoints to overcome information insufficiency due to specular reflection (as depicted in Figure~\ref{fig:fourchallenges}-(a)). It consists of single-frame pointcloud generation and non-coherent fusion of distinct frames, as shown in Algorithm~\ref{alg:NCA}.

\subsubsection{Single-Frame Pointcloud Generation}

Given the input mmWave radar ADC data, we first perform range FFT and Doppler FFT to obtain the Range-Doppler heatmap (RDM). Then, CFAR detection~\cite{gandhi1988analysisofCFAR} is applied on the Range-Doppler map to generate valid targets. Finally, the pointcloud $\mathcal{P}_{NCA, i}$ for each frame is produced with two-dimensional angle estimation and peak detection (see Section~\ref{sec:radarModel}).

\subsubsection{Coordinate Transformation from Different Perspectives}

For each timestamp $t_i$ in the accumulated time $T = {t_0, t_1, ..., t_{N}}$, we can obtain the radar pointcloud $\mathcal{P}_{NCA, i}$, device position $p_{x,i},p_{y,i},p_{z,i}$, and orientation $q_{i}$. 
Then, the coordinate transformation function $\mathcal{P}_{NCA} = \Sigma_{i=0}^{N} Q(P_{NCA, i}, q_{i}) + [p_{x,i}, p_{y,i}, p_{z,i}]$ is applied to generate the output pointcloud $\mathcal{P}_{NCA}$. The NCA effectively addresses pointcloud sparsity caused by specular reflection through non-coherent superposition of multiple frames under different device poses. However, the limited angular resolution of mmWave radar still results in $\mathcal{P}_{NCA}$ lacking local detail features.

\subsection{Synthetic Aperture Accumulation (SAA)}

The Synthetic Aperture Accumulation (SAA) aims to synthesize larger apertures using phase information from different chirps, achieving higher angular resolution compared to NCA. However, more precise position estimation (sub-mm-level for SAA, cm-level for NCA) is necessary. The non-linear and non-uniform radar motion in indoor environments typically makes most existing SAR~\cite{gao2021mimoSAR,ulander2003SARBackProjection} algorithms ineffective.
To address this issue, we propose an improved SAA imaging algorithm that adapts the back-projection method and considers the device motion characteristic and the FoV limitation. The SAA comprises signal resampling, FoV mask generation, and MIMO-back-projection imaging algorithm.

\subsubsection{Signal Resampling}
The generic SAR imaging algorithm~\cite{SARImaging2020TIP,SARAutofocusTIP2012} is inadequate for ubiquitous indoor scenarios, where the radar frequently changes its orientation or comes to a halt during motion. To address this issue, we propose a resampling method. The method involves calculating the cumulative moving distances $[d_1, d_2, ..., d_N]$ and fitting mappings to actual timestamps $[t_1, t_2, ..., t_N]$ ($F_t$), antenna positions $[p_1, p_2, ..., p_N]$ ($F_p$), antenna angles $[q_1, q_2, ..., q_N]$ ($F_q$), and measured radar signals $[R_1, R_2, ..., R_N]$ ($F_R$). Then, we obtain interpolated values under the ideal cumulative moving distances $[d_1', d_2', ..., d_N']$ using Eq.~\ref{eq:resample}:

\begin{eqnarray}
    \left\{\begin{matrix}
    [t_1', t_2', ..., t_N'] & = & F_t([d_1', d_2', ..., d_N']) \\
    [p_1', p_2', ..., p_N'] & = & F_p([d_1', d_2', ..., d_N']) \\
    [q_1', q_2', ..., q_N'] & = & F_q([d_1', d_2', ..., d_N'])  \\
    [R_1', R_2', ..., R_N'] & = & F_R([d_1', d_2', ..., d_N'])
    \end{matrix}\right.
    \label{eq:resample}
\end{eqnarray}

\subsubsection{FOV Mask}

Since the radar antenna is usually directional with a limited FoV ~\cite{rao2017introductionTIFMCWradar}, it is necessary to apply a mask according to the FoV during the back-projection process.  The mask generation algorithm consists of three steps: (i) rotating global voxels $V$ into the antenna coordinate; (ii) converting each voxel $v(x,y,z)$ into polar coordinates $v(r,\theta,\phi)$; (iii) setting mask values to 1 if the polar coordinate range of voxel $v(r,\theta,\phi)$ is within the antenna's FOV, and 0 otherwise.

\subsubsection{MIMO - Backprojection Imaging Algorithm}

The general process of the MIMO back-projection~\cite{ulander2003SARBackProjection} algorithm is as follows: Firstly, a global coordinate system is established and 3D voxels are created based on this system. Next, the round-trip delay for a given voxel $v_t$ located at $p_m'=(x_m', y_m', z_m')$, and transmit-receive antenna pair $(Rx_i, Tx_j)$ is calculated as:
\begin{equation}
    \tau^{m}_{ij} = \frac{1}{c} \|(x_m', y_m', z_m') - p_{Tx_j}\|_2 + \frac{1}{c} \|(x_m', y_m', z_m') - p_{Rx_i}\|_2
    \label{eq:delay}
\end{equation}
Subsequently, the corresponding phase variation is given by $\phi^{m}_{ij}= exp(-j2\pi f \tau^{m}_{ij})$. Finally, phase compensation is applied to obtain the contribution of each antenna position to every voxel. The total contribution from all antennas yields the final reconstruction value of the voxel.
After obtaining the voxels, we apply a threshold to discretize the SAA output, resulting in a pointcloud $\mathcal{P}_{SAA}$ that exhibits a higher angular resolution with more high-frequency feature information compared to $\mathcal{P}_{NCA}$.

\begin{figure*}[h]
    \centering
    \includegraphics[width=0.9\linewidth]{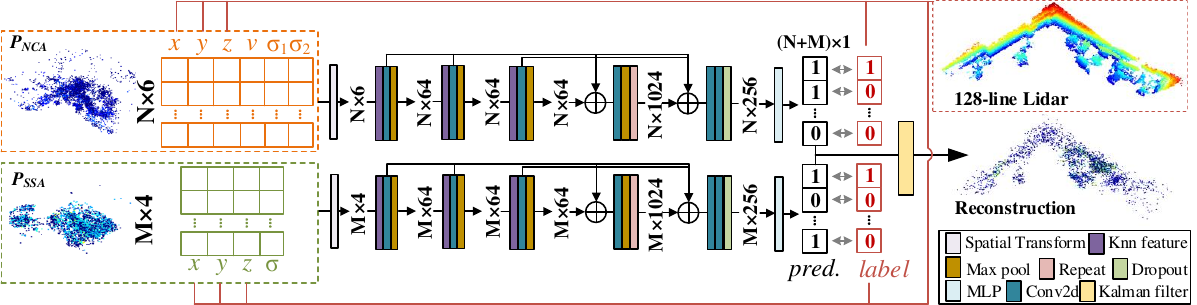}
    \caption{The architecture of RDM (Real-Denoise Multiframe) network. By leveraging LiDAR data, RDM innovatively transforms the challenging pointcloud reconstruction problem into an easily learnable point-wise binary classification task. Furthermore, RDM exploits the multi-frame accumulation capabilities of NCA and SAA, enabling the network to capture more local and global features. Inspired by DGCNN~\cite{wang2019DGCNN}, we adopt this network structure that evaluates whether each point is subject to multipath interference and noise.}
    \label{fig:RDM}
\end{figure*}

\subsection{Real-Denoise Multiframe (RDM) Network}

\subsubsection{Motivation of RDM Network}
Through the signal processing components, Non-Coherent Accumulation (NCA) and Synthetic Aperture Accumulation (SAA), we have obtained mmWave radar pointcloud that contain sufficient information and improved resolution. However, there are still a considerable amount of interference and noise in the generated pointcloud. Due to the ill-posedness nature of these interferences, traditional signal processing methods struggle to effectively eliminate them. A viable alternative is to use deep neural networks for pointcloud denoising. However, due to the substantial differences between mmWave radar pointcloud and LiDAR pointcloud, existing methods often face severe overfitting, significantly limiting the practicality of these methods. 
The proposed RDM incorporates novel designs addressing the two issues that lead to overfitting:

\vspace{0.3em}
\noindent \textbf{Demanding learning objectives.} 
The goal of mmWave radar pointcloud networks, including RDM, is to accomplish $\mathcal{P}_{hq}=\Theta(F(Radar_{ADC}))$ (Eq.~\ref{eq:deepPCD}). Most deep-learning approaches~\cite{prabhakara2022RadarHD,cai2023millipcd,sun2021deeppoint} use lidar pointcloud as supervision, assuming $\mathcal{P}_{lidar}\approx \mathcal{P}_{hq}$. However, the wavelength of mmWave radar (4mm) is approximately 2580 times larger than that of LiDAR (1550nm), resulting in significant differences in their object surface reflection properties (i.e., BRDF/RCS). Consequently, using $\mathcal{P}_{lidar}$ as supervision forces the network to learn a highly ill-posed mapping process between mmWave radar data and lidar pointcloud, instead of focusing on eliminating multipath and noise, which will lead to poor generalization capabilities~\cite{fei2022comprehensiveReviewOfPointCloudCompletion}. Figure~\ref{fig:fourchallenges}-(b) demonstrates a significant disparity between the pointcloud generated by LiDAR and mmWave radar.

\vspace{0.1em}
\noindent \underline{\textit{RDM's Solution}}: Rather than using $\mathcal{P}_{lidar}$ as supervision, RDM annotates the input pointcloud $\mathcal{P}_{NCA}$ and $\mathcal{P}_{SAA}$ based on $\mathcal{P}_{lidar}$. If a lidar point exists in the neighborhood of an input radar point, it will be labeled as a ``clean point"; otherwise, it will be regarded as a ``noise point." Therefore, RDM only performs a point-wise binary classification task (Real-Denoise) instead of pointcloud reconstruction during training and testing, significantly reducing ill-posedness and improving generalization. 

\vspace{0.3em}
\noindent \textbf{Insufficient input features.} In Eq.~\ref{eq:deepPCD}, overfitting may occur if the network $\Theta$ is forced to learn complex features from $Radar_{ADC}$, which contains only limited target information due to the specular reflection nature of mmWave. This lack of comprehensive information in a single $Radar_{ADC}$ frame makes it challenging for the network to learn intrinsic local and global features necessary for high generalization reconstruction.

\vspace{0.1em}
\noindent \underline{\textit{RDM's Solution}}: RDM incorporates multi-frame accumulation of $\mathcal{P}_{NCA}$ and multi-chirp accumulation of $\mathcal{P}_{SAA}$ to provide sufficient scene information. This approach enables the network to learn local and global features, improving performance while maintaining real-time capability through causal system design (only using data prior to time $t$).

\subsubsection{Architecture of RDM Network} In what follows, we present the network structure and loss function design based on the above analysis

\vspace{0.4em}

\noindent \textbf{Input and output.} The network input consists of pointcloud $\mathcal{P}_{NCA}$ generated by the NCA module and $\mathcal{P}_{SAA}$ produced by the SAA module. $\mathcal{P}_{NCA}$ includes 6 feature dimensions: spatial coordinates $x, y, z$, velocity $v$, and signal-to-noise-ration (SNR) in the Range-Doppler Map ($\sigma_1$) and AOA estimation ($\sigma_2$); $\mathcal{P}_{SAA}$ contains 4 feature dimensions: spatial coordinates $x, y, z$, and intensity ($\sigma$) during SAR imaging. Each mmWave radar's labels are assigned based on the presence of LiDAR points in their neighborhood:
\begin{equation}
    \label{eq:label}
    \begin{split}
        & label_i = \left\{\begin{matrix}
        1, & \text{if } \exists p_{lidar} \in N(p_i) \\
        0, & \text{otherwise}
        \end{matrix}\right. \\
    \end{split}
\end{equation}
where $N(p_i)$ is the neighborhood of $p_i$. The network output target is the point-wise $label_i$, indicating whether each point is considered to be valid or noise. The final pointcloud $\mathcal{P}_{RDM}$, after removing multipath interference and noise, is $\mathcal{P}[label=1]$.

\vspace{0.4em}
\noindent \textbf{Network.} 
Inspired by DGCNN~\cite{wang2019DGCNN}, we design the network structure as shown in Figure~\ref{fig:RDM}. Essentially, the RDM network performs a binary classification semantic segmentation task after the aforementioned transformation. The spatial transformation ensures rotational and permutation invariance of the pointcloud, while the KNN layer~\cite{zhang2017learningkforKNN} and convolutional layers ensure the extraction of local and global features of the pointcloud, respectively. 
Similar to DGCNN, the RDM recalculates the nearest neighbors for each point in the feature space at every layer during each iteration. Specifically, the RDM network comprises 12 convolutional layers, 6 KNN layers, and 8 max pooling layers. A Dropout layer is employed to mitigate overfitting, while the MLP layer performs the ultimate binary classification task.
Finally, a Kalman filter~\cite{li2015kalmanfilter} is used to post-process the network's predictions, enabling better fusion of multi-frame accumulation information from $NCA$ and $SAA$.

\subsubsection{Loss Function} \label{sec:RMDloss}
Since the output of the RDM network is a binary pointcloud classification, the commonly used binary cross-entropy loss function is a candidate. However, as most points in the input pointcloud $\mathcal{P}_{NCA}$ and $\mathcal{P}_{SAA}$ are noise and multipath, the RDM network faces an imbalance of samples and difficult cases. Directly using the binary cross-entropy loss function would lead to unsatisfactory network performance. Therefore, we adopt Focal Loss~\cite{lin2017focalloss} as the final loss function:
\begin{equation}
    \label{eq:FL}
    \begin{split}
        FL(p_t) = -\alpha_t(1-p_t)^\gamma \log(p_t)
    \end{split}
\end{equation}
where $p_t$ is the predicted probability. $(1-p_t)^\gamma$ is the modulating factor, allowing the network to focus more on difficult samples. $\alpha_t$ is $\alpha$ for ``clean points" and $1-\alpha$ for ``noise points", alleviating the class imbalance issue. In this work, we set $\alpha=0.25$ and $\gamma=2$.

The most crucial design of the RDM network is converting the challenging pointcloud reconstruction task into an easily learned point-wise binary classification task. This enables the network to focus on learning the characteristics of multipath interference and noise, rather than impractical LiDAR features. Additionally, the multi-frame accumulation greatly increases the amount of information in the input $\mathcal{P}_{NCA}$ and $\mathcal{P}_{SAA}$, which facilitates the network's learning of more local and global features.

\section{Dataset: RadarEyes}
The validation process of DREAM-PCD is hindered by the absence of high-quality mmWave radar pointcloud datasets for indoor scenarios, particularly those that can simultaneously support the full implementation of DREAM-PCD (requiring both non-coherent multi-frame data and coherent SAR data). To address this challenge, we introduce the largest indoor mmWave radar dataset, RadarEyes, containing over 1,000,000 temporally-aligned frames of mmWave radar, LiDAR, and camera data. Our experiments also reveal that the data quality of RadarEyes significantly surpasses existing  indoor mmWave radar datasets(Figure~\ref{fig:generalization}-(c)). 

\subsection{Sensor Platform} \label{sec:sensorPlatform}

\begin{figure*}[h]
    \centering
    \includegraphics[width=0.9\linewidth]{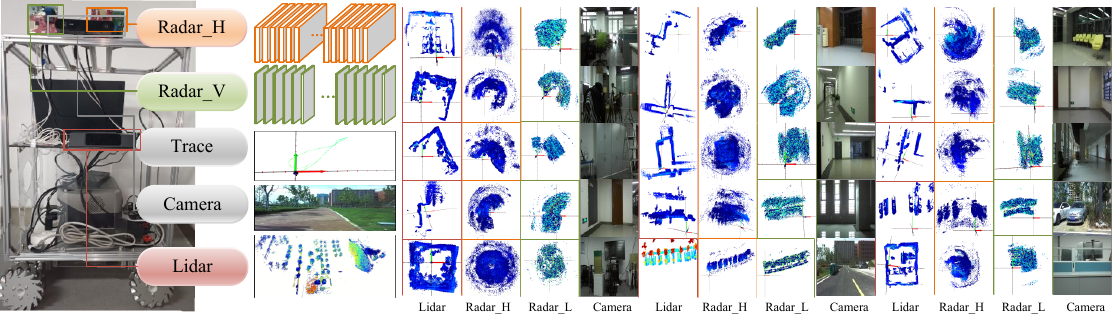}
    \caption{The data collection platform and a sample of the RadarEyes dataset. Left: The RadarEyes data collection platform. Right: A sample of the RadarEyes dataset, including LiDAR pointcloud, raw pointcloud from the two mmWave radars (without DREAM-PCD reconstruction), and camera images.}
    \label{fig:datasetSystem}
\end{figure*}

Figure~\ref{fig:datasetSystem} illustrates our sensor platform, built on a remotely controllable vehicle~\cite{xie2022embracingSensys} and equipped with four sensors:

\vspace{0.2em}
\noindent \textbf{Two orthogonally placed mmWave radars:} Both of them are Texas Instruments AWR1843BOOST-EVM paired with a DCA1000-EVM for raw data capture. Each of the radar boards has 3 transmitting antennas and 4 receiving antennas, which operate at 77-81 GHz. It is noteworthy that RadarEyes contains raw ADC data and the source code for loading and processing the ADC data, which can greatly facilitate the related research.

\vspace{0.2em}
\noindent \textbf{An 128-beam LiDAR:} LS-128 S2 1550nm~\cite{LeishenLidar} automotive-grade hybrid solid-state LiDAR.

\vspace{0.2em}
\noindent \textbf{Camera and IMU:} ZED 2i Camera~\cite{noauthorzednodate}, an integrated simultaneous localization and mapping device featuring an optical odometry and IMU. The frame rate is set to 30fps, and the localization accuracy is under $1\%$ closed-loop drift under intended use conditions.

\begin{table}
  \caption{Parameter configuration of radar and lidar}
  \label{tab:freq}
  \resizebox{\linewidth}{!}{
    \begin{tabular}{lccc}
      \toprule
                        &  Horizontal Radar   &  Vertical Radar   &  Lidar  \\
      \midrule
      Modulation        &  FMCW                 &  FMCW               &  Pulse  \\
      Framerate         &  10                  &  100                &  10     \\
      Frequency         &  $77.70-78.90GHz$     &  $79.1-80.70GHz$    &  $193THz$ \\
      TX antennas       &  3                    &  3                  &  -      \\
      RX antennas       &  4                    &  4                  &  -      \\
      Range Resolution  &  0.12             &  0.09375            &  Lidar  \\
      Max Range         &  15.99                &  12.0               &  Lidar  \\
      Azimuth Resolution&  $11.3^\circ$         &  $45^\circ$         &  $0.09^\circ$  \\
      Azimuth FOV       &  $\pm 50^\circ @6dB$  &  $\pm 20^\circ @6dB$ & $\pm 60^\circ$\\
      Elevation Resolution  &  $45^\circ$           &  $11.3^\circ$         &  $0.1^\circ$  \\
      Elevation FOV         &  $\pm 20^\circ @6dB$  &  $\pm 50^\circ @6dB$  &  $\pm 12.5^\circ$  \\
      Velocity Resolution   &  $0.12 m/s$           &  $0.55 m/s$           &  -  \\
      Max Velocity          &  $7.63 m/s$           &  $4.47 m/s$           &  -  \\
    \bottomrule
    \end{tabular}
  }
\end{table}

\subsection{Unique Advantages of RadarEyes}

In recent years, numerous mmWave radar datasets have been introduced, such as nuScenes~\cite{caesar2020nuscenes}, RadarScenes~\cite{schumann2021radarscenes}, TJ4DRadSet~\cite{TJ4DRadSet}, Oxford~\cite{barnes2020oxford}, ColorRadar~\cite{kramer2022coloradar}, and Ghost~\cite{kraus2021radarGhostDataset}. Compared to these existing datasets, RadarEyes provides the following advantages (as shown in Table~\ref{tab:compareDatasets}):

\vspace{0.2em}
\noindent \textbf{(1) Large Size.} RadarEyes contains a large amount of raw ADC radar data from various scenes.

\noindent \textit{Quantity.} Indoor perception datasets are rare compared to those for autonomous driving~\cite{zhang2021RADDet,wang2021rodnet,lim2021Radical}. Previously, Coloradar~\cite{kramer2022coloradar} was the largest indoor dataset, containing 8,122 seconds with 121,830 frames. In contrast, RadarEyes includes 120,000 frames of non-coherent mmWave radar data and 1,200,000 frames of coherent radar data, making it the largest indoor mmWave radar pointcloud dataset to date.

\noindent \textit{Diversity.} We collected data from 300 indoor scenes and 10 outdoor scenes. Indoor environments include classrooms, laboratories, conference rooms, corridors, lecture halls, and lobbies. Outdoor scenarios encompass roads, parking lots, and buildings. Detailed scene descriptions will be available on our project webpage. 

\vspace{0.2em}
\noindent \textbf{(2) Multimodal and Comprehensive.} 
RadarEyes contains rich and high-quality multimodal data, including multiple types of aligned radar data for the first time.

\noindent \textit{Rich multimodal information.} Most existing datasets only provide one modality, either LiDAR or camera, as ground truth. RadarEyes includes both 128-beam LiDAR and camera data, facilitating research on various multimodal and cross-modal perception algorithms ~\cite{lu2020milliegoSensys}. 

\noindent \textit{Both incoherent and coherent.} RadarEyes is the first mmWave radar dataset to simultaneously include incoherent~\cite{prabhakara2022RadarHD} and coherent data~\cite{gao2021mimoSAR,qian20203millipoint}, enabling future research on algorithms that utilize both types of data.

\noindent \textit{Orthogonally placed single-chip radars.} RadarEyes innovatively employs two orthogonally placed single-chip radars for pointcloud generation. Despite using two single-chip radars, the cost remains considerably lower than dual-chip cascaded radar solutions, and the elevation resolution is significantly improved.

It is worth noting that the combination of one horizontal and one vertical single-chip radar, along with non-coherent and coherent multi-frame accumulation, is a promising and feasible option for indoor/near-field radar 3D imaging. The DREAM-PCD reconstruction framework proposed in Section~\ref{sec:framework} is a pioneering work, and we hope that RadarEyes will inspire more research based on this way.

\begin{table}[t!]
  \centering
  \caption{Comparison of RadarEyes with other existing mmWave radar datasets.}
  \resizebox{\linewidth}{!}{
    \begin{tabular}{llllllll}
    \toprule
    Datasets & ADC & 3D GT & Indoor & Size  & Elevation & SAR & No FI \\
    \midrule
    nuScenes~\cite{caesar2020nuscenes} & \textcolor{red}{\ding{55}} & \textcolor{darkgreen}{\ding{51}} & \textcolor{red}{\ding{55}} & 20,000s@17Hz & NA(90$^\circ$) & \textcolor{red}{\ding{55}} & \textcolor{red}{\ding{55}} \\
    RadarScenes~\cite{schumann2021radarscenes} & \textcolor{red}{\ding{55}} & \textcolor{red}{\ding{55}} & \textcolor{red}{\ding{55}} & 15,480s@17Hz & NA(90$^\circ$) & \textcolor{red}{\ding{55}} & \textcolor{red}{\ding{55}} \\
    RaDICaL~\cite{lim2021Radical} & \textcolor{darkgreen}{\ding{51}} & \textcolor{red}{\ding{55}} & \textcolor{darkgreen}{\ding{51}} &  13,129@30Hz & 57.29$^\circ$ & \textcolor{red}{\ding{55}} & \textcolor{darkgreen}{\ding{51}} \\
    Zendar~\cite{mostajabi2020Zendar} & \textcolor{darkgreen}{\ding{51}} & \textcolor{darkgreen}{\ding{51}} & \textcolor{red}{\ding{55}} & $\sim$2,700s & NA(90$^\circ$) & \textcolor{darkgreen}{\ding{51}} & \textcolor{darkgreen}{\ding{51}} \\
    Oxford~\cite{barnes2020oxford} & \textcolor{red}{\ding{55}} & \textcolor{darkgreen}{\ding{51}} & \textcolor{red}{\ding{55}} & 60,000s@4Hz & NA(90$^\circ$) & \textcolor{red}{\ding{55}} & \textcolor{red}{\ding{55}} \\
    ColoRadar~\cite{kramer2022coloradar} & \textcolor{darkgreen}{\ding{51}} & \textcolor{darkgreen}{\ding{51}} & \textcolor{darkgreen}{\ding{51}} & 8,122*2@(10+5)Hz & 28.7$^\circ$ & \textcolor{red}{\ding{55}} & \textcolor{red}{\ding{55}} \\
    \midrule
    \textbf{RadarEyes(Ours)} & \textcolor{darkgreen}{\ding{51}} & \textcolor{darkgreen}{\ding{51}} & \textcolor{darkgreen}{\ding{51}} & 12,000s*2@(10+200)Hz & 14.32$^\circ$ & \textcolor{darkgreen}{\ding{51}} & \textcolor{darkgreen}{\ding{51}} \\
    \bottomrule
    \end{tabular}
  }
  \label{tab:compareDatasets}
\end{table}

\begin{figure}[h]
    \centering
    \includegraphics[width=0.95\linewidth]{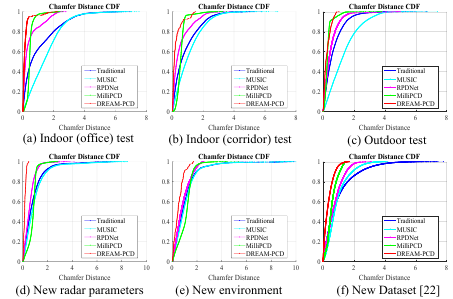}
    \caption{Chamfer Distance distributions for various baselines: (a)-(c) tested on RadarEyes; (d)-(e) tested on additional data with different parameters and environments; (f) tested on the publicly available Coloradar~\cite{kramer2022coloradar} dataset.}
    \label{fig:baselinesQualitative}
\end{figure}

\begin{figure*}[h]
    \centering
    \includegraphics[width=0.95\linewidth]{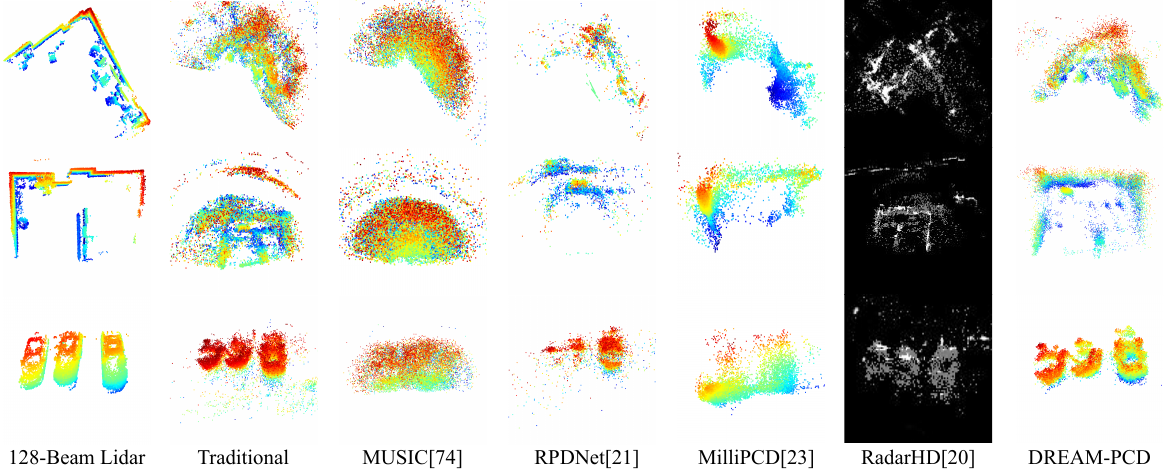}
    \caption{Qualitative comparison against baseline methods. The first two rows show results in indoor scenarios, while the third row shows results in an outdoor scenario. Note that RadarHD, proposed in ~\cite{prabhakara2022RadarHD}, is limited to generating 2D bird's-eye view (BEV) images, instead of 3D pointcloud.}
    \label{fig:baselinesQuantive}
\end{figure*}

\section{Performance Evaluation}

In this section, we first introduce the setup employed for evaluating the proposed DREAM-PCD framework, which includes evaluation metrics, baseline methods, and training details. Following that, we evaluate the performance of DREAM-PCD in terms of \textit{pointcloud reconstruction quality}, \textit{ablation study}, \textit{generalization ability}, and \textit{real-time performance}, by conducting experiments on the RadarEyes and the previous public dataset~\cite{kramer2022coloradar}. 

\subsection{Experimental Setting}

\noindent \textbf{Metrics.} Let $\mathcal{P}_{out}$ denote the pointcloud reconstruction, and $\mathcal{P}_{GT}$ represent the set of LiDAR pointcloud. The evaluation employs three metrics:

\noindent \textit{(1) Chamfer Distance (CD)}~\cite{wu2021balancedCDLoss} measures point set closeness based on nearest neighbor distances, but may not capture structural differences: 

\begin{equation}
\begin{aligned}
CD(\mathcal{P}_{out}, \mathcal{P}_{GT}) = & \frac{1}{|\mathcal{P}_{out}|}\sum_{p_{i} \in \mathcal{P}_{out}} \mathop{\min}_{p_{j} \in \mathcal{P}_{GT}} ||p_{i} - p_{j}||_{2}^{2} \\
& + \frac{1}{|\mathcal{P}_{GT}|}\sum_{p_{j} \in \mathcal{P}_{GT}} \mathop{\min}_{p_{i} \in \mathcal{P}_{out}} ||p_{i} - p_{j}||_{2}^{2}.
\end{aligned}
\end{equation}

\noindent \textit{(2) Earth Mover's Distance (EMD)}~\cite{rubner2000EMDLoss} quantifies work needed for pointcloud transformation, effectively capturing topological and geometrical differences: 

\begin{equation}
\begin{split}
EMD(\mathcal{P}_{out}, \mathcal{P}_{GT}) = & \mathop{\min}_{\phi: \mathcal{P}_{out} \rightarrow \mathcal{P}_{GT}} \\
& \frac{1}{|\mathcal{P}_{out}|} \sum_{p_{i} \in \mathcal{P}_{out}}||p_{i} - \phi(p_{i})||_{2}^{2}
\end{split}
\end{equation}

\noindent \textit{(3) F-Score}~\cite{tatarchenko2019singleFScore} assesses reconstruction quality via precision and recall: 
\begin{equation}
\textit{F-Score}(\mathcal{P}_{out}, \mathcal{P}_{GT}) = \frac{2 \times \mathcal{P}_{tr} \times R_{tr}}{\mathcal{P}_{tr} + R_{tr}}
\end{equation}
where $\mathcal{P}_{tr}$ and $R_{tr}$ denote precision and recall under threshold $tr$.

These metrics consider noise distribution (CD)~\cite{wu2021balancedCDLoss}, global distribution (EMD)~\cite{rubner2000EMDLoss} and local structure (F-Score)~\cite{tatarchenko2019singleFScore}, providing a comprehensive performance evaluation.

\begin{table}[t!]
  \centering
    \caption{Quantitative evaluation of DREAM-PCD, including comparisons against baselines and ablation experiments in indoor and outdoor scenarios.}
    \resizebox{\linewidth}{!}{
    \begin{tabular}{p{6em}|rrrrrr}
    \toprule
    \multirow{2}[4]{*}{Methods} & \multicolumn{3}{p{13.63em}}{Testset (Indoor)} & \multicolumn{3}{p{12.57em}}{Testset (outdoor)} \\
\cmidrule{2-7}    \multicolumn{1}{c|}{} & \multicolumn{1}{p{4.19em}}{CD $\downarrow$} & \multicolumn{1}{p{4.19em}}{EMD  $\downarrow$} & \multicolumn{1}{p{5.25em}}{F-Score  $\uparrow$} & \multicolumn{1}{p{4.19em}}{CD $\downarrow$} & \multicolumn{1}{p{4.19em}}{EMD $\downarrow$} & \multicolumn{1}{p{4.19em}}{F-Score $\uparrow$} \\
    \midrule
    Traditional & 5.12  & 1.30  & 0.15  & 1.49  & 0.74  & 0.13  \\
    MUSIC~\cite{schmidt1986multipleMUSIC} & 4.80  & 1.24  & 0.09  & 3.05  & 1.13  & 0.04  \\
    RPDNet~\cite{cheng2022RPDNet} & 0.63  & 0.28  & 0.14  & 1.88  & 2.14  & 0.10  \\
    MilliPCD~\cite{cai2023millipcd} & 5.78  & 0.63  & 0.05  & 3.52  & 10.52  & 0.03  \\
    \midrule
    w./o. NCA & 1.61  & 0.34  & 0.15  & 5.64  & 5.68  & 0.03  \\
    w./o. SAA & 0.36  & 0.31  & 0.16  & 2.83  & 3.51  & 0.06  \\
    w./o. RDM & 46.75  & 2.93  & 0.14  & 1.40  & \textbf{0.73}  & 0.10  \\
    \midrule
    DREAM-PCD & \textbf{0.27} & \textbf{0.26} & \textbf{0.18} & \textbf{1.36} & 1.22  & \textbf{0.18} \\
    \midrule
    \end{tabular}
    }
    \label{tab:baselinesQualitative}
\end{table}

\noindent \textbf{Baselines.} Our proposed framework is benchmarked against five baseline methods, including a traditional pointcloud generation approach~\cite{kramer2022coloradar}, the spectral estimation-based MUSIC algorithm~\cite{schmidt1986multipleMUSIC}, and recent deep learning methods RPDNet~\cite{cheng2022RPDNet}, MilliPCD~\cite{cai2023millipcd}, and RadarHD~\cite{prabhakara2022RadarHD}. As RadarHD is limited to 2D reconstruction, we perform a qualitative comparison using the bird's-eye view (BEV) representation of $\mathcal{P}_{out}$.

\noindent \textbf{Training details.} 
The RDM network, implemented in PyTorch, is trained on a single NVIDIA Tesla A100 GPU. We employ the SGDR~\cite{loshchilov2016sgdr} and early stopping~\cite{ying2019overviewEarlyStop} techniques for optimization. As the RadarEyes dataset is the only one that supports both NCA and SAA modules, we use 150 groups of data for training and the remaining 50 groups for testing. Besides, the coloradar~\cite{kramer2022coloradar} dataset is utilized to assess generalization performance.

\subsection{Pointcloud Reconstruction Experiments}

\noindent \textbf{Quantitative comparison.} Table~\ref{tab:baselinesQualitative} presents the performance of various methods in different scenarios. Our proposed framework surpasses other methods in all evaluation metrics. Notably, the improved performance in CD and EMD metrics suggests that our framework effectively mitigates multipath effects and filters noisy pointcloud. The superior F-Scores further demonstrate our framework's ability to reconstruct pointcloud with more local and global structure. Figure~\ref{fig:baselinesQualitative} (a)-(e) display the cumulative probability distribution function (CDF) plots for different CD thresholds under different test conditions. Our method has more than $80\%$ of points with a CD error below $0.3m$, compared to $60\%$ for other methods with the same error threshold.

\noindent \textbf{Qualitative comparison.} Figure~\ref{fig:baselinesQuantive} visually demonstrates the effectiveness of our proposed method in both indoor and outdoor environments. In challenging indoor settings with severe multipath, the pointcloud generated by our method significantly reduces multipath and noise interference while preserving the global structure. The fusion of horizontal and vertical radar enhances elevation plane resolution, leading to improved three-dimensional visualization, as indicated by the color distribution of points. In outdoor environments, our method also effectively removes noise, which is crucial for autonomous driving.

\begin{figure}[h]
    \centering
    \includegraphics[width=0.95\linewidth]{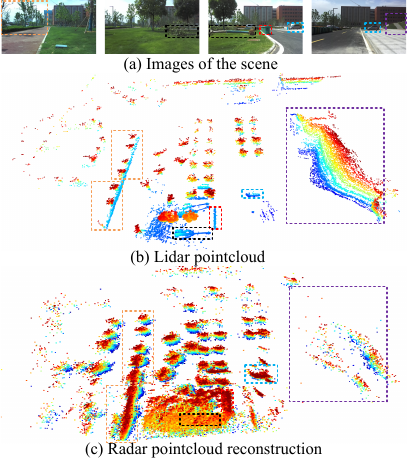}
    \caption{DREAM-PCD enables high-quality pointcloud reconstruction of large-scale scenarios using 3T4R radar.}
    \label{fig:largeScene}
\end{figure}

DREAM-PCD can generate high-resolution pointcloud for large scenes and wide-angle views due to the integration of incoherent (NCA) and coherent superposition (SAA) modules, which is a feature missing in existing methods~\cite{cai2023millipcd}. Figure~\ref{fig:largeScene} displays the pointcloud generation results of our framework for an entire square, which is unachievable by other methods.

\subsection{Ablation Study}

DREAM-PCD comprises three components: NCA, SAA, and RDM network. To evaluate their effectiveness, we perform ablation experiments by removing each module individually from the pointcloud reconstruction process (Table~\ref{tab:baselinesQualitative}). The results demonstrate that each module is essential for achieving high-quality pointcloud reconstruction. 

The NCA addresses the sparsity problem caused by specular reflection using multi-view non-coherent radar accumulation, which is particularly crucial given the unpredictable existence of strong reflectors and specular reflectance. As shown in Table~\ref{tab:baselinesQualitative}, without NCA, the Chamfer Distance increases by five times.

The SAA enhances the angular resolution in the reconstructed pointcloud, as evidenced by the noticeable decline in F-Score and CD metric without it.
Moreover, even without the SAA module, DREAM-PCD only utilizes a single low-cost 3T4R radar, yet still manages to achieve results far surpassing traditional methods. This further corroborates the effectiveness of multi-frame accumulation and the ``real-denoise'' design within DREAM-PCD.

\begin{figure}[h]
    \centering
    \includegraphics[width=0.95\linewidth]{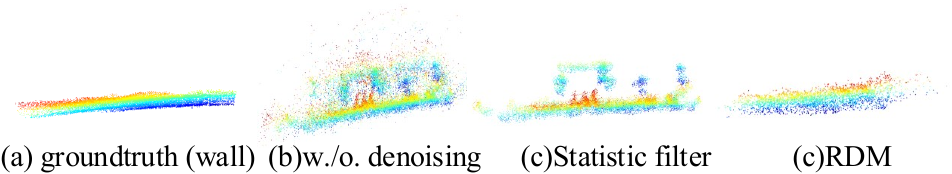}
    \caption{The ablation study of the RDM network.}
    \label{fig:ablation}
\end{figure}

Lastly, the RDM network removes multipath interference and noise with deep neural networks. Replacing the RDM network with a traditional statistical filter~\cite{balta2018faststatisticaloutlier} leads to a significant increase in ``ghost points" and noise (Figure~\ref{fig:ablation}), further emphasizing its importance (Table~\ref{tab:baselinesQualitative}). 
In outdoor scenarios, the exception in EMD is attributed to lower multipath levels and EMD's propensity to measure global pointcloud distribution, where noise may paradoxically reduce EMD in certain cases.

\begin{figure}[h]
    \centering
    \includegraphics[width=0.95\linewidth]{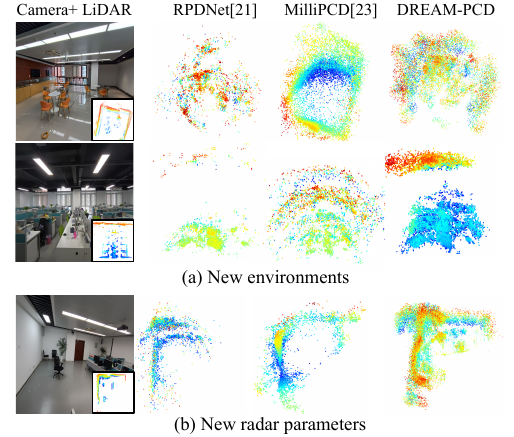}
    \caption{Generalization capability of DREAM-PCD on new radar parameters and environments.}
    \label{fig:generalization}
\end{figure}

\subsection{Generalization Experiments}

To demonstrate the generalization capability of the proposed framework, we conduct experiments with a network trained on the RadarEyes dataset and test it using data collected in new scenarios and under different radar parameters.

\noindent \textbf{New unseen environments.}
Figure~\ref{fig:generalization}-(a) illustrates different new environments which include numerous tables, chairs, and glass, and have not appeared in RadarEyes. The proposed DREAM-PCD framework demonstrates its ability to recover the overall profile without fine-tuning, while other methods failed. This indicates that the ``real-denoise" mechanism enables DREAM-PCD to learn how to extract valuable information from noisy radar data, rather than merely learning scene priors.

\begin{table*}[htbp]
    \centering
      \caption{Generalization evaluation of DREAM-PCD, including test in new radar parameters, new environments and different dataset.}
      {
          \begin{tabular}{p{6em}|p{3em}p{4em}p{4em}p{3em}p{4em}p{4em}p{3em}p{4em}p{4em}}
          \toprule
          \multirow{2}[4]{*}{Methods} & \multicolumn{3}{c}{Test in new radar para.} & \multicolumn{3}{c}{Test in new environments} & \multicolumn{3}{c}{Test in coloradar} \\
      \cmidrule{2-10}    \multicolumn{1}{c|}{} & \multicolumn{1}{c}{CD $\downarrow$} & \multicolumn{1}{c}{EMD $\downarrow$} & \multicolumn{1}{c}{F-Score $\uparrow$} & \multicolumn{1}{c}{CD $\downarrow$} & \multicolumn{1}{c}{EMD $\downarrow$} & \multicolumn{1}{c}{F-Score $\uparrow$} & \multicolumn{1}{c}{CD $\downarrow$} & \multicolumn{1}{c}{EMD $\downarrow$} & \multicolumn{1}{c}{F-Score $\uparrow$} \\
          \midrule
          Traditional & 2.64  & 0.79  & 0.10  & 2.24  & \textbf{1.17}  & 0.05  & 11.39  & 5.98  & 0.01  \\
          MUSIC~\cite{schmidt1986multipleMUSIC} & 3.68  & 0.49  & 0.04  & 2.41  & 1.22  & 0.03   & 61.33  & 7.74  & 0.00  \\
          RPDNet~\cite{cheng2022RPDNet} & 0.75  & 0.88  & 0.09  & 2.92  & 2.51  & 0.02  & 94.33  & 8.45  & 0.00  \\
          MilliPCD~\cite{cai2023millipcd} & 0.69  & 1.05  & 0.04  & 4.07  & 2.40  & 0.01  & 20.89  & 7.83  & 0.01 \\
          \midrule
          DREAM-PCD & \textbf{0.39}  & \textbf{0.40}  & \textbf{0.19}  & \textbf{2.14}  & 2.13  & \textbf{0.06}  & \textbf{10.52}  & \textbf{4.45}  & \textbf{0.02}  \\
          \bottomrule
          \end{tabular}
      }
    \label{tab:generalization}
  \end{table*}

\noindent \textbf{New Radar Parameters.}
We further assess the generalization capability of the proposed method by altering radar parameters. Specifically, we modify the radar parameters to collect test data, including changing the frequency range from 77-79GHz to 78-80.2GHz and reducing the sampling rate from 11MHz to 10MHz. The results using the original pre-trained model are shown in Figure~\ref{fig:generalization}-(b). It can be seen that the proposed method demonstrates better robustness to radar parameter changes compared to existing methods.

\noindent \textbf{New dataset.}
We evaluate the model trained on the RadarEyes dataset using the Coloradar dataset~\cite{kramer2022coloradar}. Since the Coloradar dataset does not support coherent superposition, the SAA module is not employed in this experiment. The results reveal that the proposed method outperforms traditional approaches, while existing learning-based methods fail, as illustrated in Table~\ref{tab:generalization}. Notably, the CD metric of the traditional method on Coloradar (11.39) is considerably higher than on RadarEyes (2.64), suggesting the superior quality of the proposed RadarEyes dataset.

\subsection{System Efficiency}
The proposed framework is a causal system with real-time performance in practical scenarios. Specifically, the inference speed of the framework on a DELL laptop with NVIDIA 2080 Ti GPU is approximately 0.1s per frame, allowing the processing of 10 mmWave radar frames per second. This indicates that the framework is capable of real-time pointcloud reconstruction on mmWave radar data. The live demo video in the additional materials demonstrates the system efficiency of DREAM-PCD.

\section{Limitations and Future Work}

\noindent \textbf{Multi-modal trace error correction.} The performance of the DREAM-PCD framework is affected by the accumulation of errors in the $P_{NCA}$ and $P_{SAA}$ due to position estimation errors from radar locations prior to time $t$. The trajectory information currently utilized is solely sourced from the ZED 2i camera and its internal IMU, which may potentially contain inaccuracies. One solution is to incorporate additional position constraints from mmWave radar and high-performance LiDAR through multi-sensor fusion~\cite{chghaf2022cameralidarMultiModalSLAM}, reducing localization errors and improving algorithm performance. Data collected by RadarEyes can be utilized for such research, and we will leave it as future work.

\noindent \textbf{Further shape recovery with deep models.} In DREAM-PCD, the RDM network achieves superior denoising capabilities and generalization performance by transforming the challenging reconstruction task into a simple point-wise binary classification task, resulting in high-quality pointcloud. However, the currently generated pointcloud is essentially a subset of the mmWave radar pointcloud while containing numerous scene features, making it easily achievable for further super-resolution~\cite{Dinesh2022TIPPCDVideoSR}. An improvement approach would be further enhancing the density of the pointcloud through shape recovery networks (such as GAN~\cite{li2019puGAN,sarmad2019rlGANNet}, diffusion~\cite{luo2021diffusion,lyu2021conditionalpointdiffusion}, etc.), thereby improving the visual results.

\noindent \textbf{Deployment on a drone.} The current system is implemented on a wheeled robot, limited to planar movement, and thus does not effectively improve performance in the elevation dimension. An exciting future direction is to deploy the entire system on a drone, leveraging the accumulation and denoising mechanisms of DREAM-PCD to achieve high azimuthal and elevation resolution. This could address the challenges faced by existing drone-based radar imaging systems in complex indoor environments~\cite{vergnano2022droneRadar}, with broad applications in fields such as rescue operations and reconnaissance missions.

\section{Conclusion}

In this paper, we presented DREAM-PCD, an mmWave radar pointcloud reconstruction framework for indoor scenarios, achieving superior performance through ``Real-Denoise" and multi-frame accumulation mechanisms. We also introduced RadarEyes, the largest available indoor mmWave radar dataset containing more than 1,440,000 frames collected by 4 sensors, enabling numerous applications in radar pointcloud, radar imaging, and multimodal sensing domains.
We anticipate that the DREAM-PCD framework together with the RadarEyes dataset will significantly advance learning-based mmWave pointcloud reconstruction research, and thereby enable a wide range of exciting radar sensing applications in various complex environments in the near future.

\ifCLASSOPTIONcaptionsoff
  \newpage
\fi

\bibliographystyle{IEEEtran}
\bibliography{egbib}

\end{document}